\documentclass[aps,prb,twocolumn,superscriptaddress,showpacs,amsmath,amssymb,footinbib,longbibliography]{revtex4-1}
\usepackage[english]{babel}
\usepackage{amssymb,amsbsy,amsmath}
\usepackage{graphicx}% Include figure files
\usepackage{dcolumn}% Align table columns on decimal point
\usepackage{bm}% bold math
\usepackage{subfigure}
\usepackage{color}
\usepackage{textcomp}
\usepackage[colorlinks,bookmarks=false,citecolor=blue,linkcolor=red,urlcolor=blue]{hyperref}
\newcommand{\op}[1]{%
    \fontdimen12\textfont3=2pt\fontdimen12\scriptfont3=1.4pt%
    \!\null\mathop{\vphantom{#1}\smash{#1}}\limits_{\sim}\null\!}
\newcommand{\xref}[1]{\protect\ref{#1}}
\newcommand{\figref}[1]{Fig.~\protect\ref{#1}}
\newcommand{\fmref}[1]{(\protect\ref{#1})}
\def\bra#1{\langle \, {#1} \, | \,}
\def\ket#1{\, | \, {#1} \, \rangle}
\newcommand{\braket}[2]{\langle \, {#1} \, | \, {#2} \, \rangle}

\newcommand{\SMax}{S_{\mbox{\scriptsize max}}}

\begin{document}
%
%\title{Putting translational invariance on spin systems}
\title{Combined use of translational and spin-rotational invariance for spin systems}

\author{Tjark Heitmann}
\email{tjark.heitmann@uos.de}
\affiliation{Fachbereich f\"ur Physik, Universit\"at
  Osnabr\"uck, Barbarastr. 7, D-49076 Osnabr\"uck, Germany}
\author{J\"urgen Schnack}
\email{jschnack@uni-bielefeld.de}
\affiliation{Fakult\"at f\"ur Physik, Universit\"at Bielefeld, Postfach 100131, D-33501 Bielefeld, Germany}

\date{\today}

\begin{abstract}
Exact diagonalization and other numerical studies of quantum
spin systems are notoriously limited by the exponential growth
of the Hilbert space dimension with system size. 
A common and well-known practice to reduce this increasing
computational effort is to take advantage of the translational
symmetry $C_N$ in periodic systems. 
This represents a rather simple yet elegant application of the
group theoretical symmetry projection operator technique. 
For isotropic exchange interactions, the spin-rotational
symmetry $SU(2)$ can be used, where the Hamiltonian matrix is
block-structured according to the total spin- and magnetization
quantum numbers. Rewriting the Heisenberg Hamiltonian in terms
of irreducible tensor operators allows for an efficient and
highly parallelizable implementation to calculate its matrix
elements recursively in the spin-coupling basis. When combining
both $C_N$ and $SU(2)$, mathematically, the symmetry projection
technique leads to ready-to-use formulas. However, the
evaluation of these formulas is very demanding in both
computation time and memory consumption, problems which are
said to outweigh the benefits of the symmetry reduced matrix
shape. We show a way to minimize the computational effort for
selected systems and present the largest numerically accessible
cases. 
\end{abstract}


\maketitle

%%%%%%%%%%%%%%%%%%%%%%%%%%%%%%%%%%%%%%%%%%%%%%%%%%%%%%%%%%%%%%%%%%%%%%%%
\section{Introduction}

A typical system that possesses both spin-rotational as well as
translational symmetry is a Heisenberg 
spin ring
\cite{Bet:ZP31,TDP:JACS94,SGJ:PRB96,WFP:ACIE97,Xia:PRB98,MaS:PRL98,BSS:JMMM00,Wal:PRB01,Waldmann:EPL02,WGC:PRL03,PKB:MPR03,AGC:PRB03,EnL:PRB06,WSC:PRL09,BCS:PRB09,DWD:PRB10,BTP:PNAS12,UNM:PRB12,GBC:PRB16}
which models,  e.g., certain 
magnetic molecules or chains with the following Hamiltonian,
where periodic boundary conditions are applied,
%--------------------------------------------------------
\begin{eqnarray}
\label{E-I-1}
\op{H}
&=&
-2 
J\;
\sum_{i=1}^N\;
\op{\vec{s}}_i \cdot \op{\vec{s}}_{i+1}
\ ,\quad
\op{\vec{s}}_{N+1}\equiv\op{\vec{s}}_{1}
\ .
\end{eqnarray}
%--------------------------------------------------------
The dot-product between the spin vector operators ensures spin
rotational symmetry, since dot-products do not change upon
simultaneous rotations of both vectors. The same value $J$ of
interactions between adjacent neighbors gives rise to
translational invariance, since the spin ring can be collectively
moved by one spacing without changing the Hamiltonian.

Both symmetries can be employed for various purposes. One is of
course the perception of fundamental properties without even
evaluating the energy spectrum: the energy eigenvalues form
multiplets, i.e. total spin $S$ and its magnetic quantum number
$M$ are good quantum numbers. The same holds for the momentum
quantum number $k=0, \dots N-1$, that also explains certain
degeneracies, namely between $k$ and
$N-k$.\cite{Kar94,BSS:JMMM00,Wal:PRB00,BSS:JMMM00:B,Schnack:PRB00,BSS:EPJB03} 
Together with the notion of bipartiteness these quantum numbers
can be assigned to, for instance, the ground state, again without
diagonalizing the
Hamiltonian.\cite{Mar:PRS55,LSM:AP61,LiM:JMP62}

The other application is the reduction of dimensionality
when diagonalizing the Hamiltonian. This is achieved by
block-structuring the Hamiltonian matrix according to the
available quantum numbers, or in the language of group theory,
the available irreducible representations. This powerful tool,
that is heavily used in exact diagonalization studies, is the
topic of this investigation. In order to guide the reader to the
achievements and problems of combining full spin-rotational
symmetry with translational symmetry, we present important
precursors first.

For spin problems, where at least the total magnetization $M$ is
a good quantum number, i.e. $[\op{H},\op{S}^z]=0$, one can
subdivide the full Hilbert space ${\mathcal H}$ into the 
direct sum of all eigenspaces ${\mathcal H}(M)$ of
$\op{S}^z$
%--------------------------------------------------------
\begin{eqnarray}
\label{E-I-2}
{\mathcal H}
=
\bigoplus_{M=-\SMax}^{+\SMax}\;
{\mathcal H}(M)
\ .
\end{eqnarray}
%--------------------------------------------------------
This is easily achieved by sorting the product basis states
$\ket{m_1, m_2, \dots, m_N}$ according to their total magnetic
quantum number $M=\sum_{i=1}^N m_i$, which yields basis states
$\ket{m_1, m_2, \dots, m_N; M}$ in each orthogonal subspace
${\mathcal H}(M)$.\cite{SHS:JCP07} This scheme is employed in
many popular codes for exact and approximate diagonalization, as
for instance, by means of Density Matrix Renormalization Group
(DMRG), compare, e.g., the ALPS package.\cite{ALPS:JMMM07,BCE:JSMTE11}

To marry the $\op{S}^z$-symmetry with translational symmetry is again
rather easy since the irreducible representations of the
translations can be constructed analytically starting from
states $\ket{m_1, m_2, \dots, m_N; M}$. If $\op{T}$ denotes a
translation of the chain by one site, i.e. the generating group
operation of the translation group $C_N$, then
%--------------------------------------------------------
\begin{eqnarray}
\label{E-I-3}
\ket{m_1, m_2, \dots, m_N; M, k}
\propto&&
\\
\sum_{\nu=0}^{N-1}
\left(e^{i 2\pi k/N} \op{T}\right)^\nu
&&
\ket{m_1, m_2, \dots, m_N; M}
\nonumber
\end{eqnarray}
%--------------------------------------------------------
is both an eigenstate of $\op{S}^z$ and $\op{T}$ with
eigenvalues $M$ and $\exp(-i 2\pi k/N)$, respectively,
$k=0,\dots N-1$ being
the shift quantum number (lattice momentum). 
After considering that cyclic permutations of $m_1, m_2, \dots,
m_N$ yield the same $\ket{m_1, m_2, \dots, m_N; M, k}$ and that
some patterns $m_1, m_2, \dots, m_N$ with additional symmetry do
contribute only to certain $k$, one can set
up a very straight forward generation of the basis states in the
subspaces ${\mathcal H}(M,k)$, whose dimensions are about $1/N$th
of the respective dimensions of ${\mathcal
  H}(M)$.\cite{BSS:JMMM00,SZP:JPI96,RSH:PRB04,RLM:PRB08,RiS:EPJB10}
This scheme is also used in many programs, 
among which \verb§spinpack§ is a freely available
one.\cite{spin:256}
Application in DMRG seems to be restricted
since matrix-product states are constructed according to
positions of spins, therefore each state breaks translational
invariance. Nevertheless, very recently ideas have been
developed how to incorporate translational symmetry into
DMRG.\cite{ZVH:PRB18}

Then, what is the problem with the combination of full
spin-rotational symmetry and translational symmetry?

The paper is organized as follows. In Section \ref{sec-2} we
recapitulate how spin rotational and translational symmetry can
be applied simultaneously and discuss the numerical
implications. Thereafter in
Section~\ref{sec-3} we present some of the largest numerically
exact calculations for spin rings followed by a discussion in
Section~\ref{sec-4}.

%%%%%%%%%%%%%%%%%%%%%%%%%%%%%%%%%%%%%%%%%%%%%%%%%%%%%%%%%%%%%%%%%%%%%%%%
\section{Spin-rotational and translational symmetry}
\label{sec-2}

The major
obstacle when combining spin-rotational and translational
symmetry is given by the fact, that a translated eigenstate of
$\op{\vec{S}}^2$ in general does not belong to the same basis set as the
original state, in contrast to the basis $\left\{\ket{m_1, m_2,
  \dots, m_N; M}\right\}$, where translations yield just another
member of the same basis set. In order to understand this
better, we quickly repeat how spin-rotational symmetry -- $SU(2)$
-- can be realized. This is done by means of spin coupling
according to some arbitrary coupling scheme. The basis states
%--------------------------------------------------------
\begin{eqnarray}
\label{E-S-1}
\ket{s_1, s_2, S_{12}, s_3, S_{123}, \dots, s_N, S, M}
\end{eqnarray}
%--------------------------------------------------------
are e.g. generated by sequential coupling of spins along the
chain. They are by construction eigenstates of $\op{\vec{S}}^2$
and $\op{S}^z$. If the Hamiltonian is then written in terms of
irreducible tensor operators that are connected to compound
tensors according to the same coupling scheme, matrix elements
of the Hamiltonian can be easily evaluated by recursive
decoupling. A detailed description of this powerful method can
be found in references
\onlinecite{GaP:GCI93,BCC:IC99,BeG:EPR,Tsu:group_theory,Tsu:ICA08,BBO:PRB07,ScS:PRB09,ScS:IRPC10}.   
The computer program \verb§MAGPACK§, that completely
diagonalizes the Heisenberg Hamiltonian using $SU(2)$ symmetry,
is freely available.\cite{BCC:JCC99} Also for DMRG $SU(2)$ codes
have been
developed.\cite{MCG:EPL02,McC:JSM07,FKM:JPA11,Alv:CPC12,Wei:AP12,ZFK:JSM16}
In other fields such as nuclear 
physics this method was also adapted to model finite Fermi
systems such as nuclei \cite{DuP:RPP04} as was the case for
Hubbard models, where one can actually exploit two $SU(2)$
symmetries.\cite{AZH:PRB88,Zha:IJMPB91,LNN:PRB98,Sch:AP02}
Solutions for models with $SU(N)$ symmetry work along similar
lines.\cite{MCG:EPL02,TRS:PRB12,NaM:PRL14,NaM:PRB16,WNM:PRB17}

\begin{figure}[ht!]
\centering
\includegraphics*[clip,height=30mm]{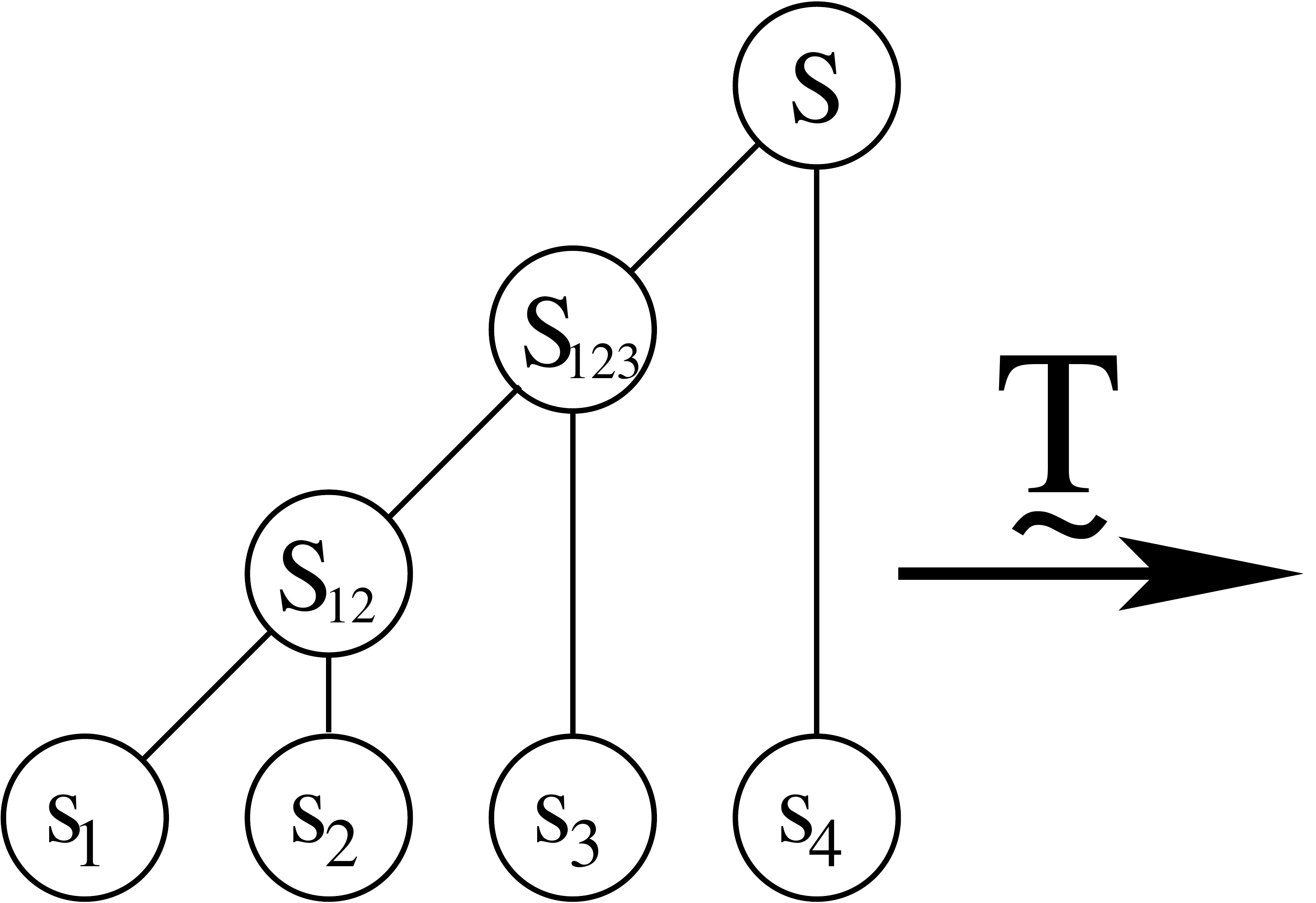}
\includegraphics*[clip,height=30mm]{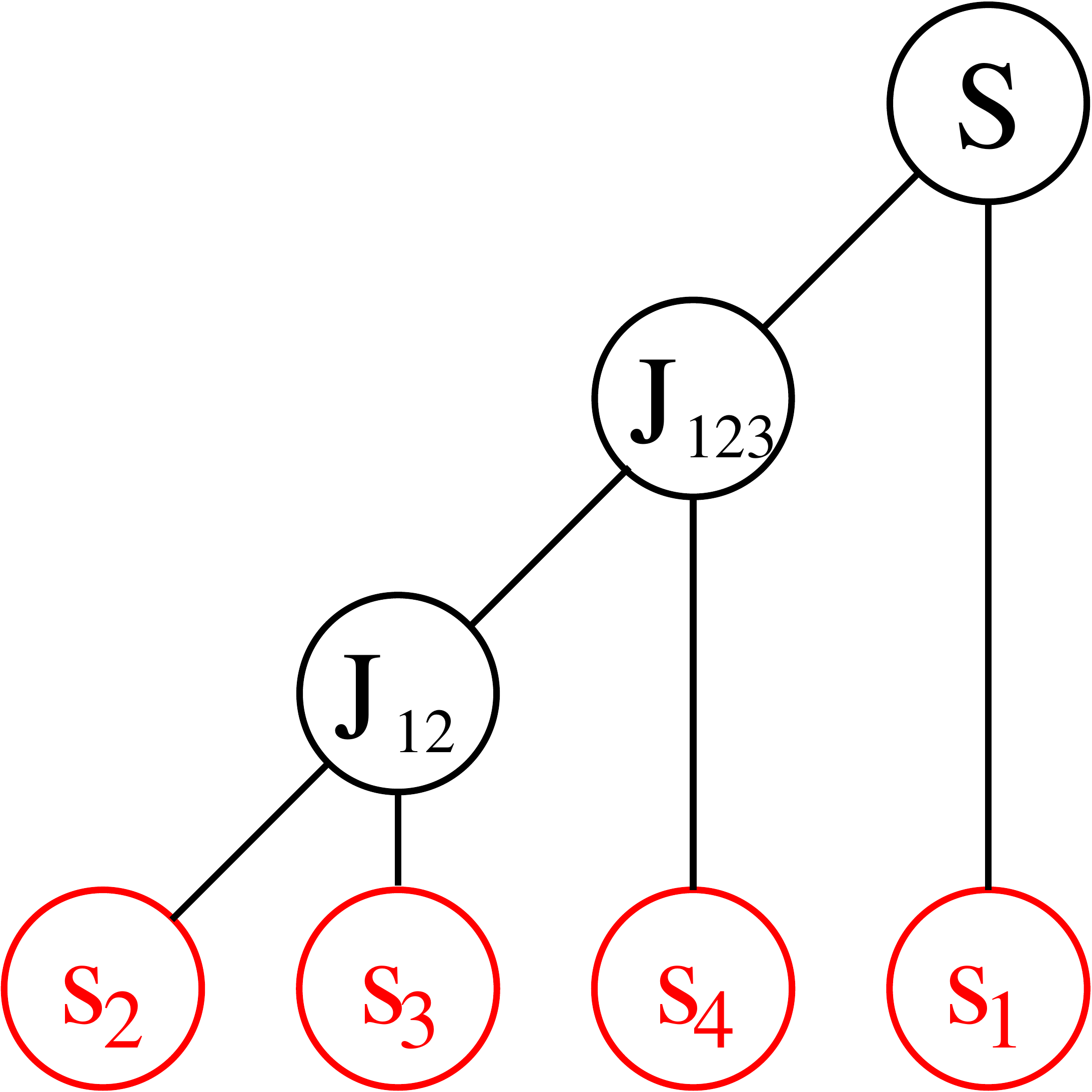}
\caption{(Color online) Coupling schemes can be represented as
  coupling trees. The original sequential coupling (l.h.s.) is
  transformed into a sequential coupling that starts with the
  spin at position 2 (r.h.s.). The intermediate spins are labeled with a
  different letter to denote the different coupling scheme,
  although they acquire the original value, i.e. $J_{12}=S_{12}$.}
\label{transspin-f-A}
\end{figure}

The construction of a new basis that is in addition an eigenbasis of the
translation operator $\op{T}$ involves the projection
operator already introduced in \fmref{E-I-3},
%--------------------------------------------------------
\begin{eqnarray}
\label{E-S-2}
\ket{\alpha, S, M, k}
\propto
\sum_{\nu=0}^{N-1}
\left(e^{i 2\pi k/N} \op{T}\right)^\nu
&&
\ket{\alpha, S, M}
\ .
\end{eqnarray}
%--------------------------------------------------------
Here $\alpha$ is now a short-hand notation for the full coupling
scheme $s_1, s_2, S_{12}, s_3, S_{123}, \dots, s_N$. To be used
as a basis, the states $\ket{\alpha, S, M, k}$ still need to be
orthonormalized. The application of $\op{T}$ in \eqref{E-S-2}
generates a plethora of new
states that belong to \emph{different} coupling schemes, i.e. to
 \emph{different} basis sets. Figure~\xref{transspin-f-A}
 demonstrates the action of $\op{T}$ on a coupling
 scheme of a ring of four spins. 
The translation of all spins by one unit modifies the whole
coupling scheme, which is in stark contrast to the
 action on product states $\ket{m_1, m_2, \dots, m_N}$,
 where only a new member of the same basis set is produced.

%===================    figure   =================================
%===================    figure  =================================

In order to evaluate matrix elements of the Hamiltonian each 
state $\op{T}^\nu \ket{\alpha, S, M}$ has to be represented in the
original basis, i.e. 
%--------------------------------------------------------
\begin{eqnarray}
\label{E-S-3}
\op{T}^\nu \ket{\alpha, S, M}
=&&
\nonumber
\\
\sum_{\alpha^\prime}
\ket{\alpha^\prime, S, M}
&&
\bra{\alpha^\prime, S, M}\op{T}^\nu \ket{\alpha, S, M}
\ .
\end{eqnarray}
%--------------------------------------------------------
Thanks to symmetry this needs to be done only for e.g. $M=S$,
but it nevertheless involves a huge number of so-called 
recoupling coefficients $\bra{\alpha', S, M=S}\op{T}^\nu
\ket{\alpha, S, M=S}$. Graph-theoretical methods can be used to
evaluate these
coefficients,\cite{FPV:CPC95,FPV:CPC97,ScS:PRB09,ScS:IRPC10}
which contain Wigner-6J symbols, phase factors, square roots as
well as possibly summations over additional indices. The
composition of these coefficients is crucial for the
computational costs of not only their calculation but also the
time and memory efficiency of the whole basis symmetrization. 
Defining an equivalence relation 
%--------------------------------------------------------
\begin{eqnarray}
\label{E-2-6}
&&\ket{\alpha',S,M}\cong\ket{\alpha,S,M}
\nonumber
\\
\Leftrightarrow\quad\exists  \nu &:& \bra{\alpha',S,M}\op{T}^{\nu}\ket{\alpha,S,M}\neq 0
\end{eqnarray}
%--------------------------------------------------------
enables to distinguish orthogonal sets of projected states which
can be orthonormalized separately. The number and size of these
sets is closely related to the complexity of the recoupling
coefficients, where simple coefficients lead to many small
sets. 
In the worst case, where all states are equivalent,
orthonormalization becomes cumbersome and one needs to store an
order of  
$\left[\text{dim}({\mathcal H}(S,M=S))\right]^2/N$
basis coefficients. This prevents a general use even for
relatively small systems.

The complexity of the recoupling coefficients depends on several
circumstances, in particular the used point group and the
employed coupling scheme.\cite{ScS:PRB09,ScS:IRPC10} The
relevant question is therefore, 
whether coupling schemes exist that are substantially less
demanding than others. In an earlier publication it could be
shown that if one chooses compatible point groups and coupling
schemes, only phase factors appear in the recoupling
coefficients.\cite{Wal:PRB00} Since especially low-symmetry
groups such as $D_2$ 
or $D_4$ often allow for the construction of an appropriate
coupling scheme,\cite{DGP:IC93,BOS:TMP06,SBO:JPA07} we wonder
whether also the group of translations $C_N$ can be combined
with a clever coupling scheme.

%===================    figure   =================================
\begin{figure}[ht!]
\centering
\includegraphics*[clip,height=25mm]{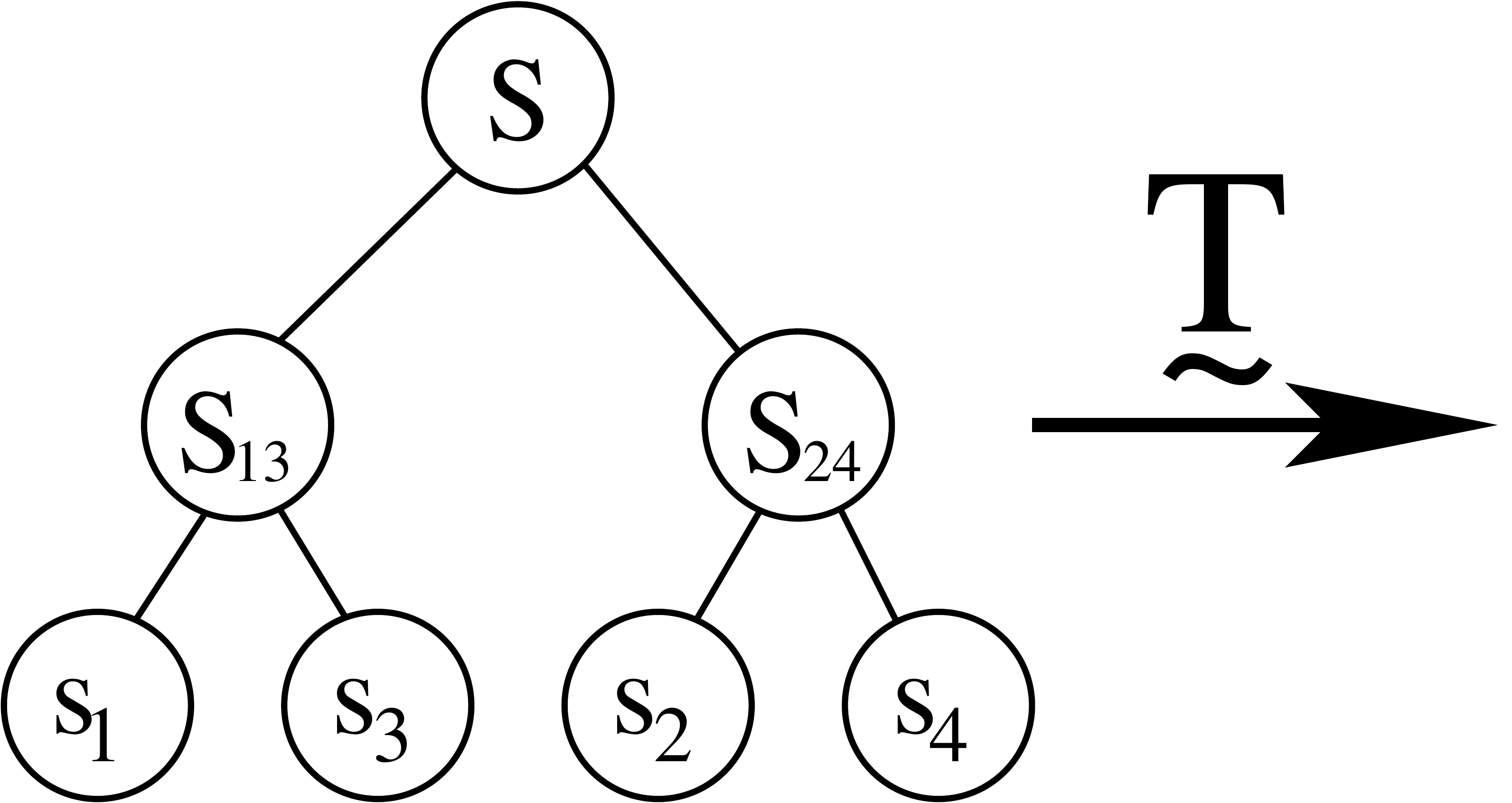}
\includegraphics*[clip,height=25mm]{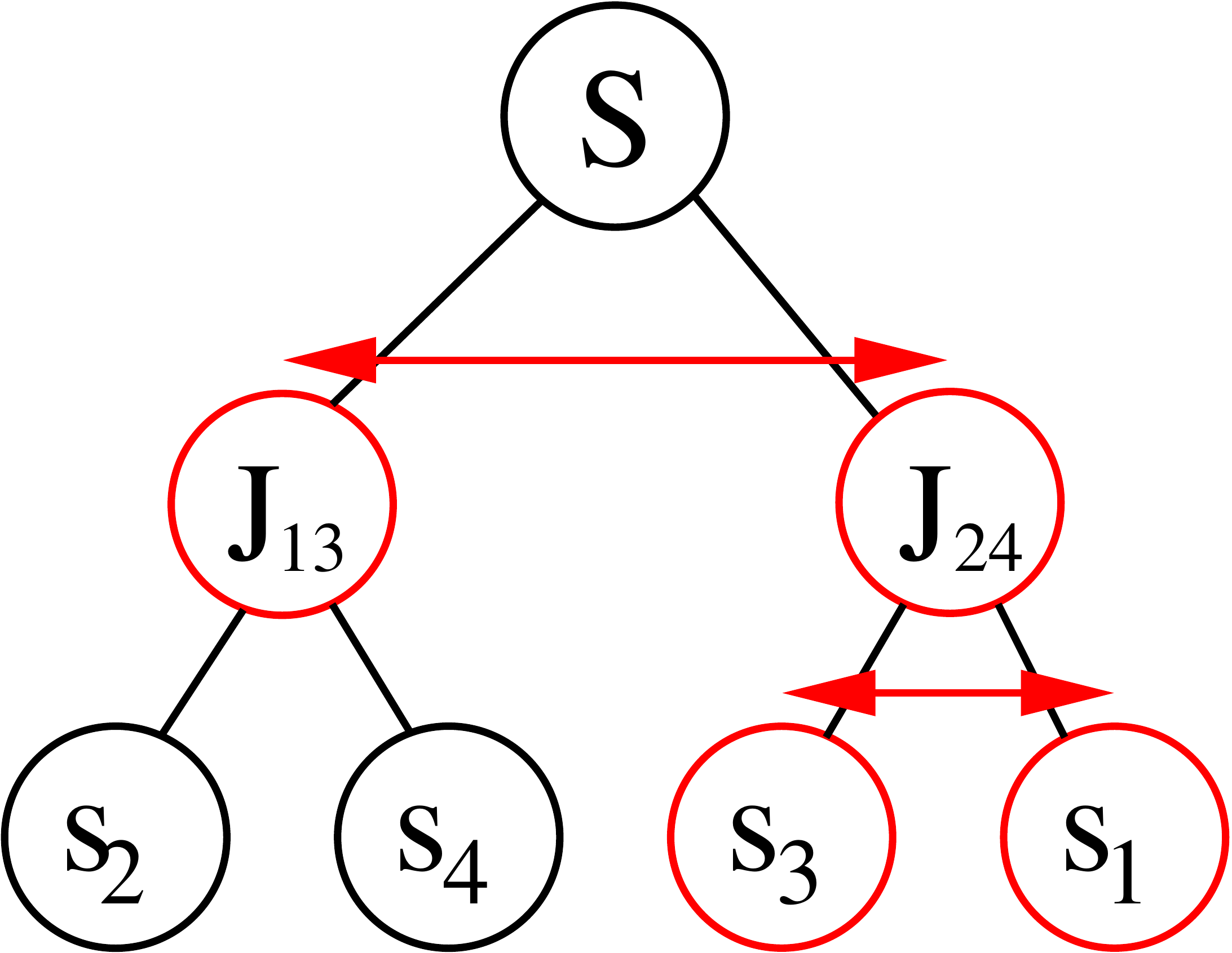}
\caption{(Color online) Optimal coupling scheme for chain
lengths of $N=2^n$ (l.h.s.). The translated scheme (r.h.s.) can
be transformed back into the old coupling scheme by spin
exchange operations on the coupling graph, leading to a very
simple recoupling coefficient.}
\label{transspin-f-B}
\end{figure}
%===================    figure  =================================

The mentioned graph-theoretical methods
\cite{FPV:CPC97,FPV:CPC95,ScS:PRB09,ScS:IRPC10} help to
understand what one is looking for: recoupling coefficients
without summations over additional indices and with as few as
possible Wigner symbols and square roots. The ultimate goal --
no sums, no symbols, no square roots -- can be achieved for chain
lengths of $N=2^n, n=2, 3, 4, \dots$. Then the recoupling
coefficients can be 
evaluated in the graph-theoretical framework by spin exchange
processes as depicted in \figref{transspin-f-B}. Such processes
generate only a phase, as for example in $\braket{s_1 s_2 S}{s_2
s_1 S}=(-1)^{S-s_1-s_2}$. For the example shown on the r.h.s. of  
\figref{transspin-f-B} this yields ($M=S$ omitted)
%--------------------------------------------------------
\begin{eqnarray}
\label{E-S-4}
&&\braket{s_1, s_3, S_{13}, s_2, s_4, S_{24}, S}{s_2, s_4,
  J_{13}, s_3, s_1, J_{24}, S}
\nonumber
\\
&&=
(-1)^{J_{24}-s_3-s_1}
(-1)^{S-J_{13}-J_{24}}
\delta_{S_{13}J_{24}}
\delta_{S_{24}J_{13}}
\nonumber
\\
&&=
(-1)^{S-J_{13}-s_3-s_1}
\delta_{S_{13}J_{24}}
\delta_{S_{24}J_{13}}
\ .
\end{eqnarray}
%--------------------------------------------------------

%===================    figure   =================================
\begin{figure}[ht!]
\centering
\includegraphics*[clip,width=0.9\columnwidth]{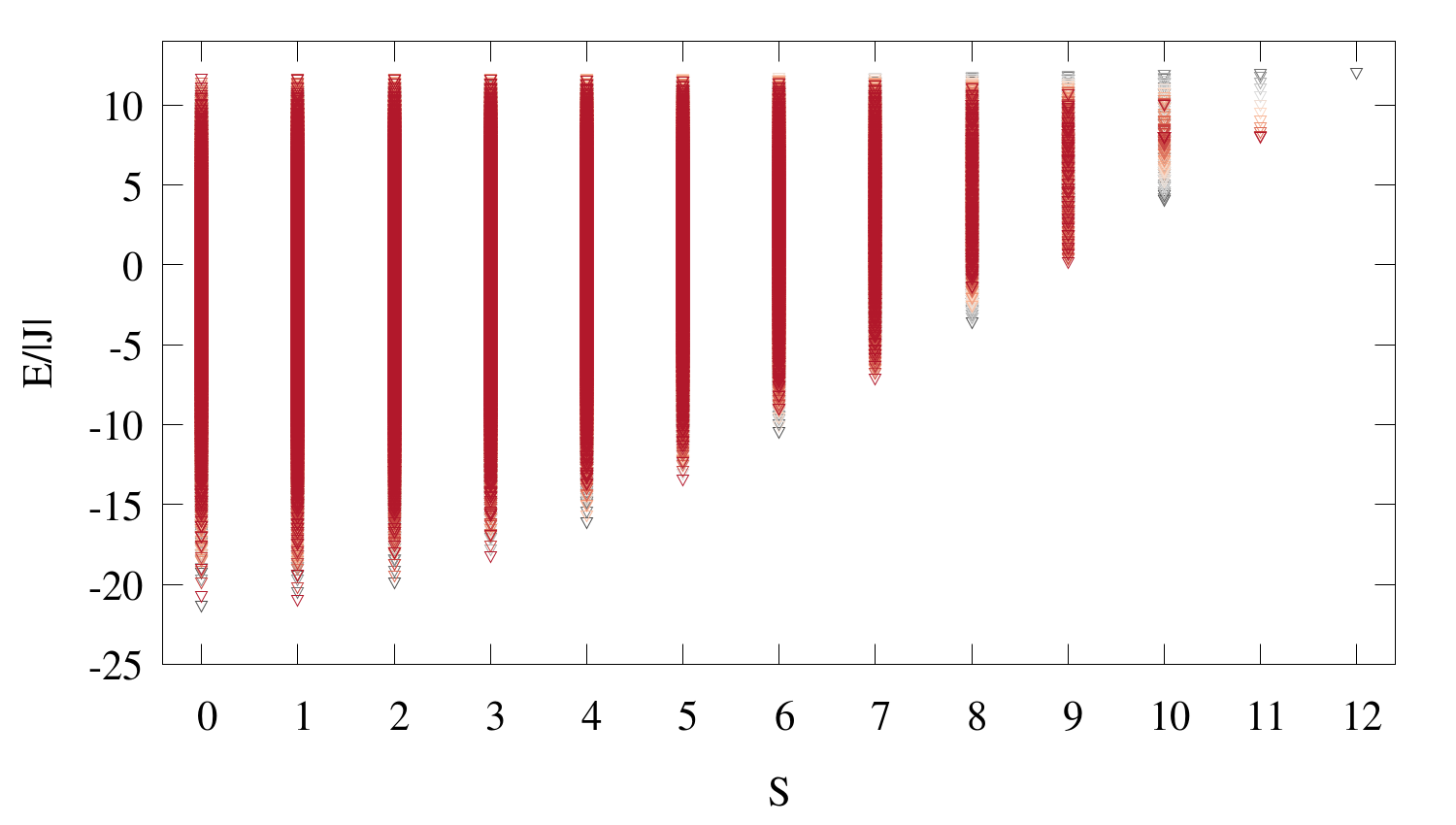}

\includegraphics*[clip,width=0.9\columnwidth]{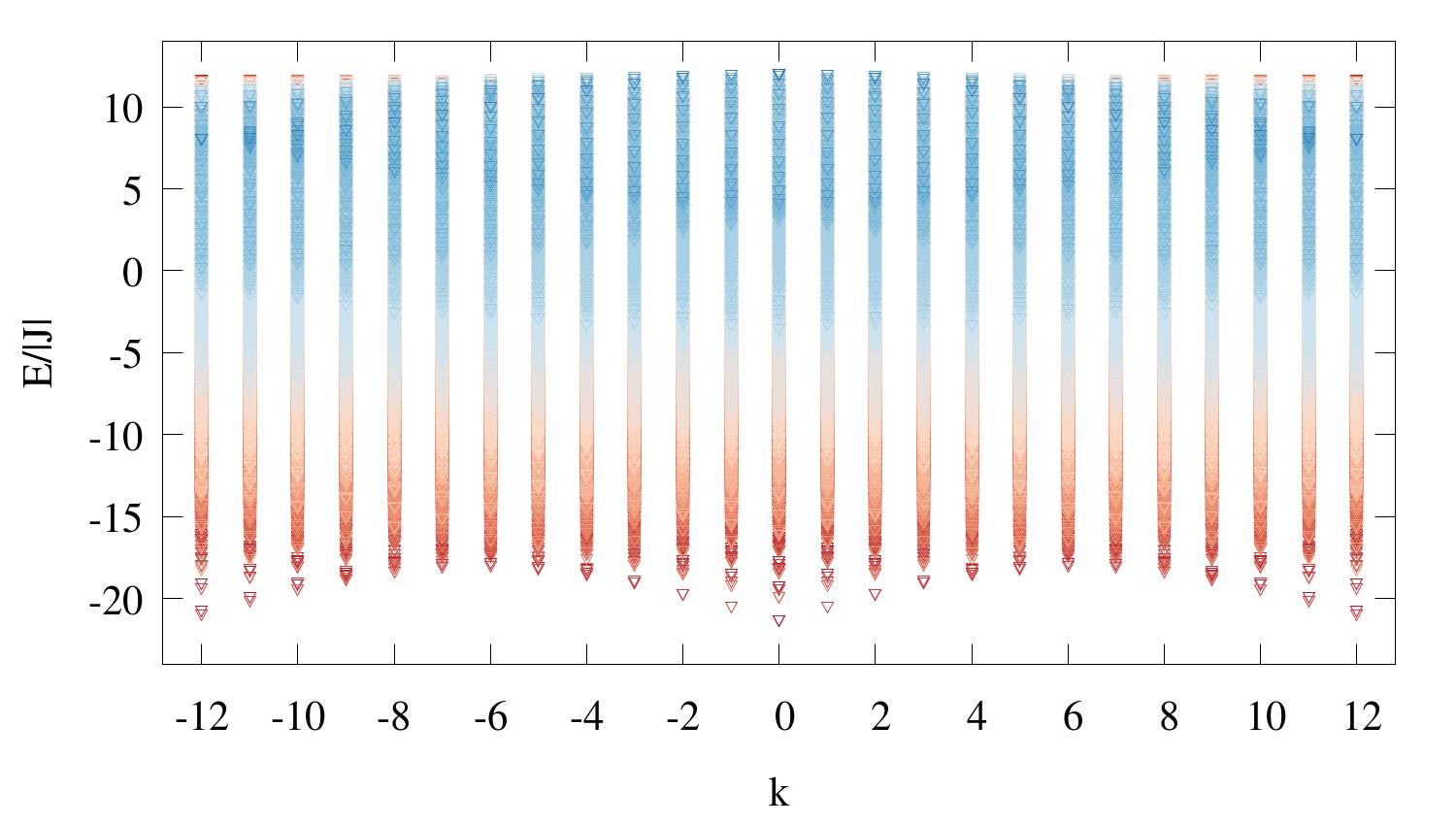}

\includegraphics*[clip,width=0.9\columnwidth]{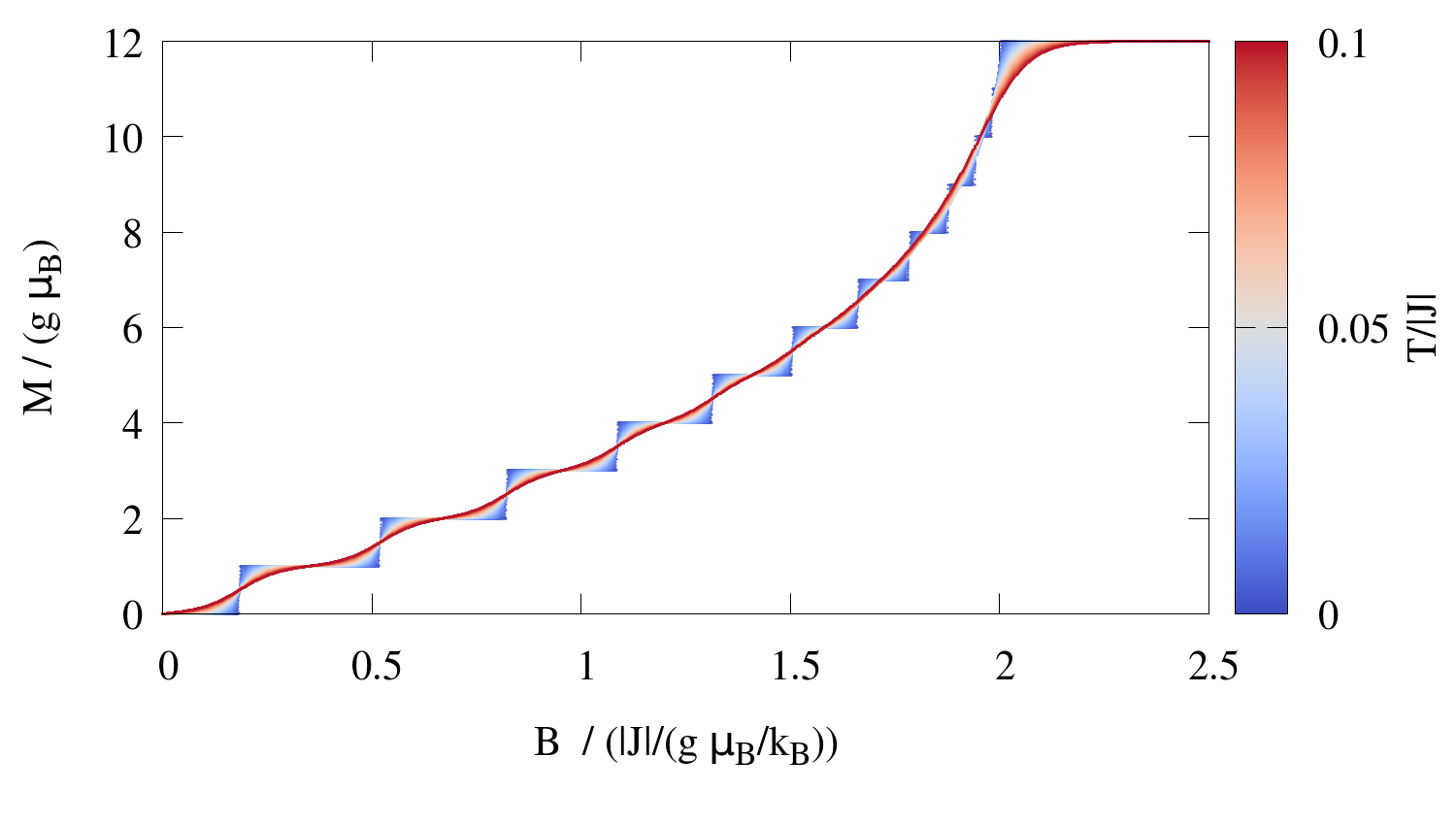}

\includegraphics*[clip,width=0.9\columnwidth]{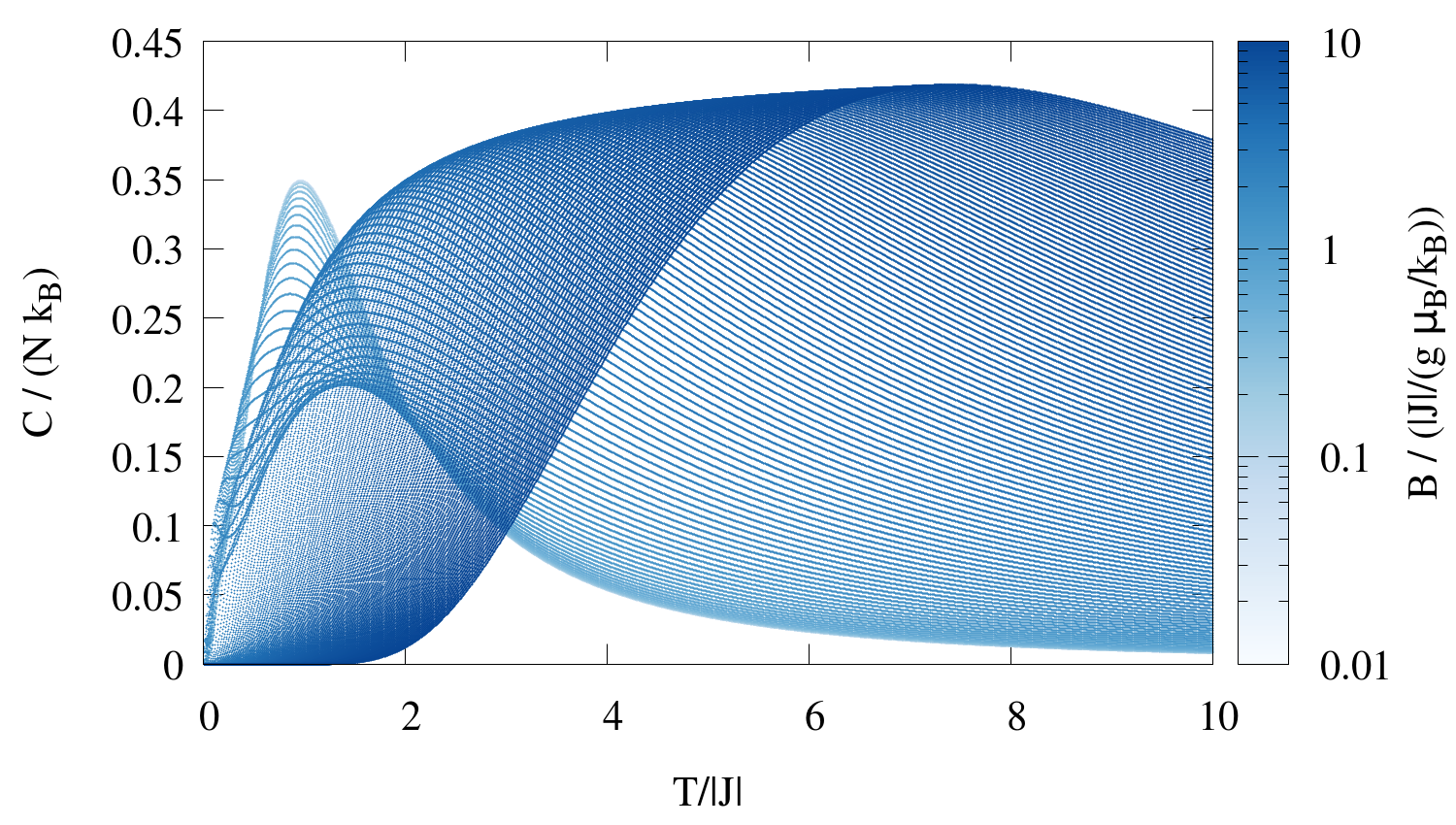}
\caption{(Color online) Spectra and observables for an
  antiferromagnetic Heisenberg ring with $N=24$, $s=1/2$:
energy spectrum vs total spin $S$, the same spectrum but now vs $k$,
the magnetization vs the applied field $B$ for various
temperatures as well as the specific heat vs temperature $T$ for
various external fields (top to bottom).}
\label{transspin-f-C}
\end{figure}
%===================    figure  =================================

For chain lengths that are not powers of two, it turns out that a
universal coupling strategy is to „prime factorize“ the coupling
scheme, i.e. the chain length. $N=6$ for instance would be
coded as $2\cdot 3$, and 
so on. The recoupling coefficients contain more and more Wigner
symbols as well as square roots the larger the prime factors $p_i$
are. The maximum number of symbols per coefficient is given by   
%--------------------------------------------------------
\begin{eqnarray}
\label{E-2-5}
\text{N}_\text{Wigner-6J}
&=&
\sum\limits_{i=1}^{\text{N}_\text{primes}}(p_i-2)\cdot\sum\limits_{j=i+1}^{\text{N}_\text{primes}}p_j
\ .
\end{eqnarray}
%--------------------------------------------------------
$N=2^n$ fits into this scheme as the optimal case, since
only the smallest possible prime factors appear.
This finding explains why a combination of spin-rotational and translational
symmetry is not easily possible for the majority of system sizes
-- it turns into a prohibitive
numerical effort to evaluate a massive number of recoupling
coefficients.

%===================    figure   =================================
\begin{figure}[ht!]
\centering
\includegraphics*[clip,width=0.9\columnwidth]{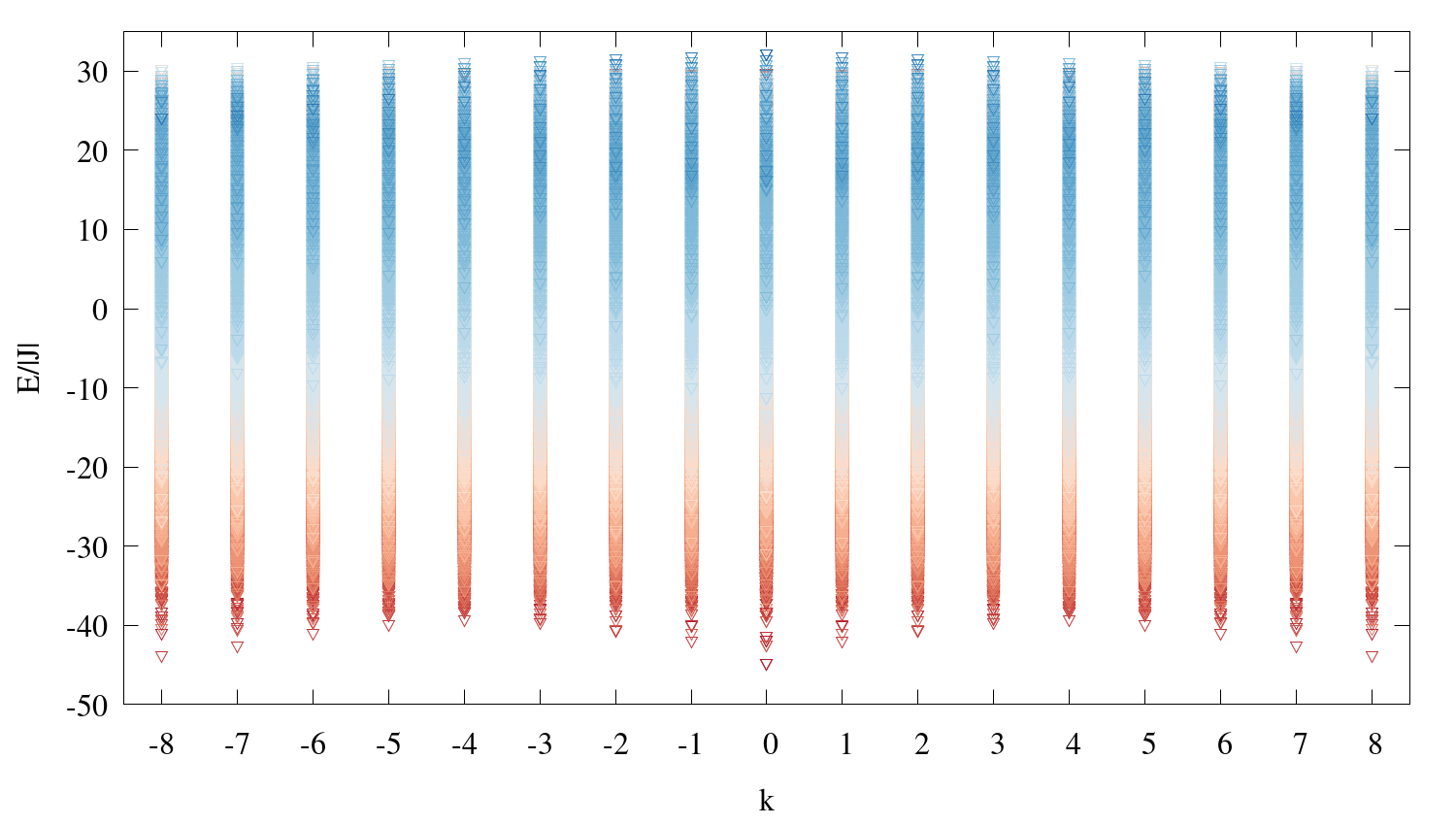}

\includegraphics*[clip,width=0.9\columnwidth]{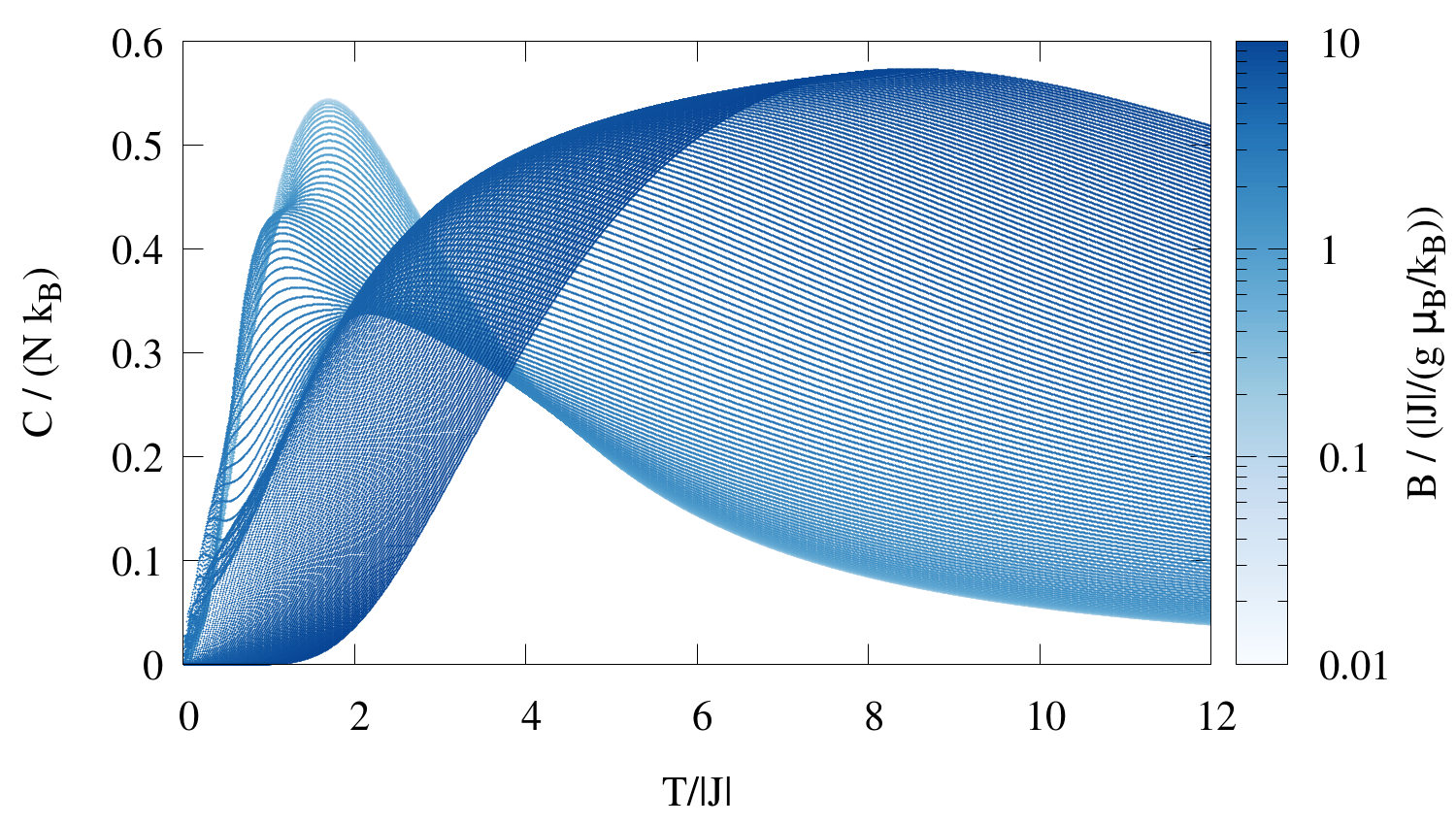}
\caption{(Color online) Spectra and observables for an
  antiferromagnetic Heisenberg ring with $N=16$, $s=1$:
energy spectrum vs $k$ as well as the specific heat vs
temperature $T$ for 
various external fields (top to bottom).}
\label{transspin-f-D}
\end{figure}
%===================    figure  =================================

%%%%%%%%%%%%%%%%%%%%%%%%%%%%%%%%%%%%%%%%%%%%%%%%%%%%%%%%%%%%%%%%%%%%%%%%
\section{Numerical results}
\label{sec-3}

Finally we would like to present some of the largest cases one
can actually solve nowadays. We choose Heisenberg spin rings with
antiferromagnetic nearest-neighbor interaction as examples.

The first example shows spectra and magnetic observables for a
spin ring with $N=24$ sites of spins $s=1/2$. The dimension of the
total Hilbert space is $\text{dim}({\mathcal H})=16,777,216$,
which can be subdivided into subspaces ${\mathcal H}(S,M=S,k)$
as outlined above. In particular, $24=2*2*2*3$. The dimension of
the largest subspace 
${\mathcal H}(S,M=S,k)$ is $27,275$; it occurs for $S=2$ and even $k\ne 0,12$.
Figure~\xref{transspin-f-C} shows from top to bottom the energy
spectrum vs total spin $S$, the same spectrum but now vs $k$,
the magnetization vs the applied field $B$ for various
temperatures as well as the specific heat vs temperature $T$ for
various external fields. The figures merely serve as visual
proofs of the feasibility of the program than as sources for
specific curves. Readers interested in the spectra or specific
functions are welcome to contact the authors.

The second example presents the results for a spin ring of $N=16$
sites of spins $s=1$. In this case the total dimension assumes a
value of $\text{dim}({\mathcal H})=43,046,721$, which
reduces to $59,143$ for the largest subspace 
${\mathcal H}(S,M=S,k)$ occurring for $S=3$ and odd $k$.
Figure~\xref{transspin-f-D} depicts the energy
spectrum vs $k$ as well as the specific heat vs temperature $T$ for
various external fields.

\begin{figure}[ht!]
\centering
\includegraphics*[clip,width=0.9\columnwidth]{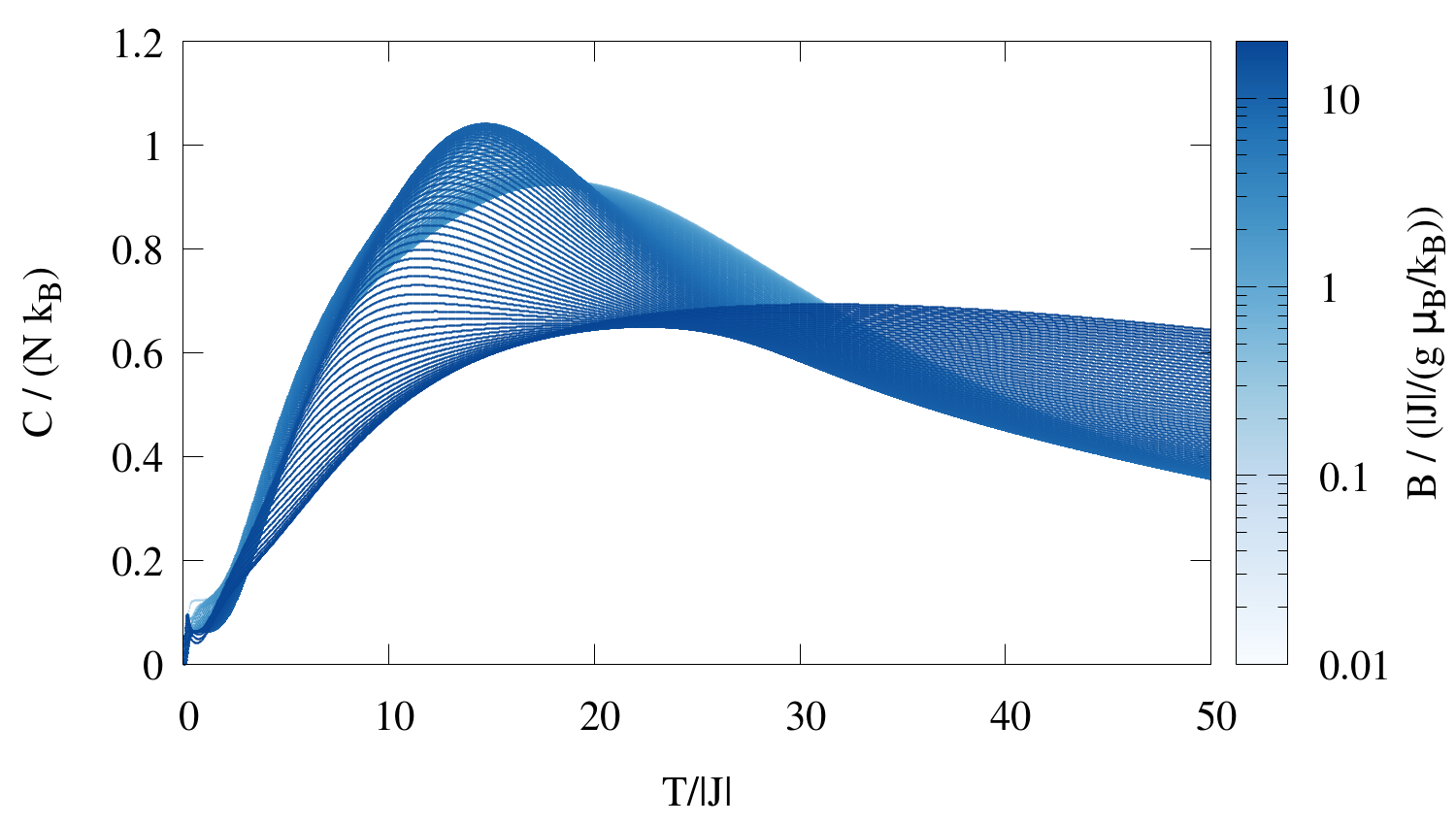}
\caption{(Color online) Specific heat vs
temperature $T$ for various external fields for an
  antiferromagnetic Heisenberg ring with $N=8$, $s=5$.}
\label{transspin-f-E}
\end{figure}

The final example of our selection deals with a fictitious spin
ring of $N=8$ spins with single-spin quantum number $s=5$. Its
main purpose is to demonstrate that the combined use of
spin-rotational as well as translational symmetry allows to
reduce the 
staggering dimension of the full Hilbert space of
$\text{dim}({\mathcal H})=214,358,881$ to a rather moderate size
of the largest subspace ${\mathcal H}(S,M=S,k)$ of
$77,970$ which occurs for $S=9$ and odd $k$.
Figure~\xref{transspin-f-E} shows the specific heat vs
temperature $T$ for various external fields calculated from all
214,358,881 levels.

%===================    figure   =================================
%===================    figure  =================================

%%%%%%%%%%%%%%%%%%%%%%%%%%%%%%%%%%%%%%%%%%%%%%%%%%%%%%%%%%%%%%%%%%%%%%%%
\section{Discussion and conclusions}
\label{sec-4}

The outlined method provides a valuable tool in cases where a
complete and numerically exact diagonalization of a large spin
system provides additional benefits compared to approximate
methods. The knowledge of exact quantum numbers such as $S$,
$M$, and $k$ provides such benefits for instance in
spectroscopic experiments as, for instance, inelastic neutron scattering
(INS), where selection rules can be
inferred.\cite{Fur:IJMPB10,BGC:NP12,FuW:RMP13}

The method also complements other existing exact methods, in
particular Bethe ansatz methods. These work for isotropic
nearest-neighbor interactions of arbitrary spin
$s$,\cite{KRS:LMP81,Tak:PLA82,Bab:NPB83,Tsv:NPB88,FYF:NPB90} but
only for certain linear combinations of 
powers of $\op{\vec{s}}_i \cdot \op{\vec{s}}_{i+1}$. The most
general isotropic nearest-neighbor interaction for 
spin $s$ is of the form $p_s(\op{\vec{s}}_i \cdot
\op{\vec{s}}_{i+1})$, where $p_s$ denotes a polynomial of degree
$2s$. For $s=1/2$ the polynomial is simply the proportional
function, which means that the Heisenberg spin-1/2 chain is
integrable by Bethe ansatz. For spin-$1$ chains the polynomial
turns out as $p_1(x) = x \pm x^2$ or $p_1(x) = x^2$, which means
that certain bilinear/biquadratic chains can be solved by
Bethe ansatz.\cite{MLE:PRB11} Generally, the Bethe ansatz is not
applicable to Heisenberg chains with only bilinear terms for $s
> \frac{1}{2}$. Here (and in many other cases) our
diagonalization scheme provides the exact spectra and
eigenfunctions, albeit for periodic chains of restricted 
lengths. 

Although the theoretical calculations appear straight
forward, we showed that in many cases a vast
number of recoupling coefficients is generated which in the
worst cases yields $\text{dim}({\mathcal H}(S,M=S))$ coefficients
for each of the $\text{dim}({\mathcal H}(S,M=S))/N$ 
states belonging to an irreducible representation $(S,M,k)$. This
renders a practical use impossible. Nevertheless, we could also
outline, for which system sizes a combined use of
spin-rotational and translational symmetry is feasible. It then 
delivers numerically exact results for both spectra as well as
observables.

Very recent numerical studies show that the range of
applicability of the method can be extended, at least
somewhat, by using $D_N$ combined with parity instead of $C_N$
symmetry.\cite{San:AIPCP10} Complex valued basis coefficients
and matrix elements can thereby be avoided at the cost of
additional symmetry operations. 
This way, a complete diagonalization of a spin ring with $N=27$
and $s=1/2$ becomes possible, for instance.

%%%%%%%%%%%%%%%%%%%%%%%%%%%%%%%%%%%%%%%%%%%%%%%%%%%%%%%%%%%%%%%%%%%%%%%%
\section*{Acknowledgment}

This work was supported by the Deutsche Forschungsgemeinschaft DFG
(314331397 (SCHN 615/23-1); 355031190 (FOR~2692); 397300368 (SCHN~615/25-1)). 
Computing time at the Leibniz Center in Garching is gratefully
acknowledged.
The authors thank Andreas Kl{\"umper} and Frank G{\"o}hmann for
helpful discussions concerning Bethe ansatz solutions and Jonas
Richter as well as Robin Heveling for carefully reading the
manuscript.

%%%%%%%%%%%%%%%%%%%%%%%%%%%%%%%%%%%%%%%%%%%%%%%%%%%%%%%%%%%%%%%%%%%%%%%%
%\bibliographystyle{/home/schnack/tex/sty/revtex4-1/revtex4-1/bibtex/bst/revtex/apsrev4-2}
%\bibliography{/home/schnack/tex/bibtex/js-own.bib,/home/schnack/tex/bibtex/js-mag.bib}

%merlin.mbs apsrev4-1.bst; modified by Andreas Honecker for title output
%Control: key (0)
%Control: author (72) initials jnrlst
%Control: editor formatted (1) identically to author
%Control: production of article title (-1) disabled
%Control: page (0) single
%Control: year (1) truncated
%Control: production of eprint (0) enabled
%

\end{document}